\documentclass{aa}

\usepackage{color}

\usepackage{graphicx}
\begin{document}
   \title{Laboratory Experiment of Checkerboard Pupil Mask Coronagraph}

   \subtitle{}

   \author{K. Enya\inst{1},
           S. Tanaka\inst{1, 2},
           L. Abe\inst{3}
           \and
           T. Nakagawa\inst{1}
          }

   \offprints{K. Enya,\\
          e-mail: enya@ir.isas.jaxa.jp}

   \institute{
        Department of Infrared Astrophysics, Institute of Space and
        Astronautical Science, \\
        Japan Aerospace Exploration Agency,
        Yoshinodai 3-1-1, Sagamihara, Kanagawa  229-8510, Japan
         \and
        Department of Physics, Graduate School of Science,\\
        University of Tokyo, Hongo 7-3-1, Bunkyo-ku, Tokyo 113-0033, Japan
         \and
        Optical  and  Infrared  Astronomy  Division \& Extra-solar Planet Project Office,
        National  Astronomical  Observatory,\\
        Osawa 2-21-2, Mitaka,  Tokyo  181-8588,  Japan
        }

   \date{Received September 15, 1996; accepted March 16, 1997}


  \abstract
   {We present the results of the first laboratory experiment
   of  checkerboard shaped pupil binary mask coronagraphs
   using visible light,
   in the context of the R\&D activities for
   future mid-infrared space missions such as the 3.5\,m
   SPICA telescope.
   }
   {The primary aim of this work is to demonstrate the
    coronagraphic performance of checkerboard masks down to
    a $10^{-6}$ peak-to-peak contrast, which is required to
    detect self-luminous extra-solar planets in the mid-infrared
   region.
   }
   {Two masks, consisting of aluminum films on a glass substrates,
were manufactured using nano-fabrication techniques with electron
beam lithography: one mask was optimized for a pupil with a 30\%
central obstruction and the other was for a pupil without obstruction.
The theoretical contrast for both masks was  $10^{-7}$ and no
adaptive optics system was employed.
   }
  { For both masks, the observed point spread functions were quite
consistent with the theoretical ones.
The average contrast measured within the dark regions was
$2.7 {\times} 10^{-7}$ and $1.1 {\times} 10^{-7}$.
   }
  {The coronagraphic performance significantly outperformed the
$10^{-6}$ requirement and almost reached the theoretical limit
determined by the mask designs. We discuss the potential
application of checkerboard masks for mid-infrared coronagraphy,
and conclude that binary masks are promising for future
high-contrast space telescopes.
  }

   \authorrunning{K. Enya, S. Tanaka, L. Abe, T. Nakagawa}
   \titlerunning{Laboratory Experiment of Checkerboard
        Pupil Mask Coronagraph}
   \maketitle
%

\section{Introduction}

Direct detection and spectroscopy of extra-solar planets
is very important to the understanding of how planetary systems
were born, how they evolve, and ultimately to find biological
signatures on these planets. After the first report unambiguously
confirming the existence of an extrasolar planet by 
Mayor \& Queloz (1995), 
more than 200 extra-solar planets
have been found to date,
mostly by detailed observation of the Doppler shift of the
spectrum of the central star (\cite{Mayor1995}), or by the
variability of the luminosity of the central star due to the
transit of
the planet (\cite{Henry2000}; \cite{Charbonneau2000}).
However, such indirect observation does not usually provide the
spectral features, luminosity and other important properties of
the planets themselves, so the direct
observation of extra-solar planets is of primary importance.
However, the enormous contrast in luminosity between the central
star and the planet, and the associated diffraction
patterns from optical instruments requires some coronagraphic
techniques to suppress, or at least attenuate them.
Typically, the contrast at visible light wavelengths is
$\sim\,10^{10}$ but is reduced to $\sim\,10^6$ in the
mid-infrared region (\cite{Burrows2004}),
which is one of the most important advantages of infrared
coronagraphy for the observation of extra-solar planets.

The Space Infrared telescope for Cosmology and
Astrophysics (SPICA) is the next generation mission
for infrared astronomy led by Japan (\cite{Nakagawa2004}).
The SPICA telescope uses on-axis Ritchey-Chretien
optics with a 3.5\,m diameter monolithic primary mirror.
The whole telescope will be cooled to 4.5\,K and
infrared observations will be made at wavelengths within the
5--200\,$\mu$m range. A coronagraphic instrument is currently
being considered for the SPICA mission
(\cite{Enya2006}; \cite{Abe2006}).
The primary target of this instrument
is the direct detection and observation of extra-solar
Jovian outer planets (typically beyond 5\,AUs).
The baseline requirements for the SPICA coronagraph are as
follows: $10^{-6}$ peak-to-peak contrast, smallest possible
inner-working-angle (IWA), and a core wavelength
within the range  $5$--$20$\,$\mu$m.

Of the various current coronagraphic methods,
coronagraphs using binary shaped  pupil  masks have some
remarkable advantages (e.g., Jacquinot \& Roizen-Dossier 1969;
Vanderbei, Spergel \& Kasdin 2003; Kasdin,
Vanderbei \& Spergel 2003).
They are relatively simple in their implementation,
essentially achromatic systems,
and robust against pointing errors of the telescope
(\cite{green2004}).
Therefore, binary masks
have been studied as the baseline for the SPICA coronagraph
development (\cite{Tanaka2006}; \cite{Enya2006}).
Our aim is to demonstrate precise mask fabrication, and
to check the performance of the masks with visible light
in the laboratory before proceeding further.

In this paper, we present the results of a laboratory
demonstration of the checkerboard mask coronagraphs in visible
light. The target contrast to be demonstrated in this
work was set to $10^{-6}$, which is required to detect
self-luminous extra-solar planets in the mid-infrared. It was shown
that the observed contrast went beyond this goal, very close
to the theoretical contrast of $10^{-7}$. Expansion toward a
mid-infrared demonstration test-bench is also discussed.


\section{Experiment}

\subsection{Pupil Mask}


Recently, Tanaka et al.\,(2006)
has studied the performance
of binary masks (circular and checkerboard types)
shaped for a pupil comprising a central obstruction and spiders,
and an asymmetric checkerboard mask design was presented.
For this work, we selected symmetric checkerboard
masks because they can be also optimized for the expected
SPICA telescope pupil geometry.
Moreover, checkerboard
masks consist of rectangular patterns which are
suitable for the fabrication process used.
The reason that asymmetric
checkerboard masks were not chosen was only in order
to make the fabrication simple; in fact,
the fabrication of the asymmetric masks
would basically present no additional
manufacturing problems.

Optimization of the mask shape was performed with
the LOQO solver presented by {\cite{ref_loqo} to find
the solution which maximizes the pupil throughput under the
constraints of a given contrast, IWA and outer working
angle (OWA). The optimization for checkerboard masks
is simply computed in 1D which provides the 2D so-called
``bar-code'' mask solution. The final 2D mask is simply the
combination of two of these masks, rotated by 90 degrees
with respect to each other. Therefore, the useful IWA for
the 2D masks is $\sqrt{2}$  times larger than the IWA defined for
the optimized bar-code mask. Note that in this paper,
all measurements are made in diagonally from
the center of the image in order to match the correct IWA
definition.

\begin{table}
  \caption{Design of the two binary checkerboard mask}\label{table01}
  \begin{center}
    \begin{tabular}{lcc}
\hline
\hline
                         & Mask\,1 & Mask\,2\\
\hline
      Type               & Symmetric   &  Symmetric \\
      Central obstruction   & 30\%       &  No obstruction \\
      IWA$^{*}$($\lambda/D$)      & 7          &  3   \\
      OWA$^{*}$($\lambda/D$)      & 16         &  30 \\
      Contrast              & 10$^{-7}$     &  10$^{-7}$ \\
      Throughput$^{**}$(\%)           &  16          &  24      \\
\hline
    \end{tabular}
  \end{center}
     \small{
      $*$ In the experiment the wavelength was $\lambda$ = 632.8\,nm, 
          and
          the diameter of the entrance pupil was $D$\,=\,2\,mm
        (the binary mask is inscribed to this pupil).

     $**$ Ratio of the transmissive area of the mask to the whole area
       for an assumed entrance pupil diameter of 2\,mm.
       }
\end{table}

\begin{figure}
\centering
  \includegraphics[width=43mm,height=43mm]{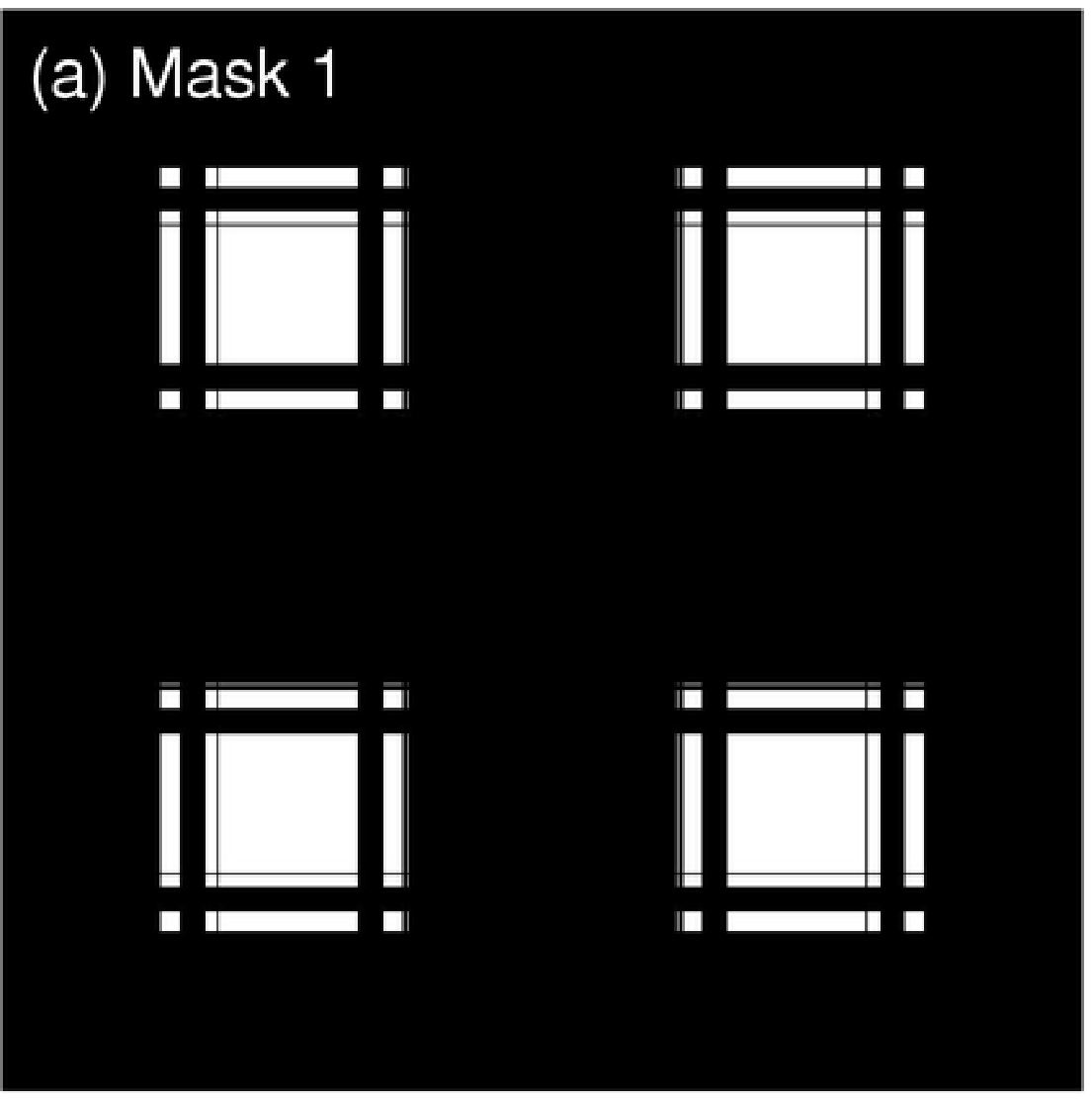}
  \includegraphics[width=43mm,height=43mm]{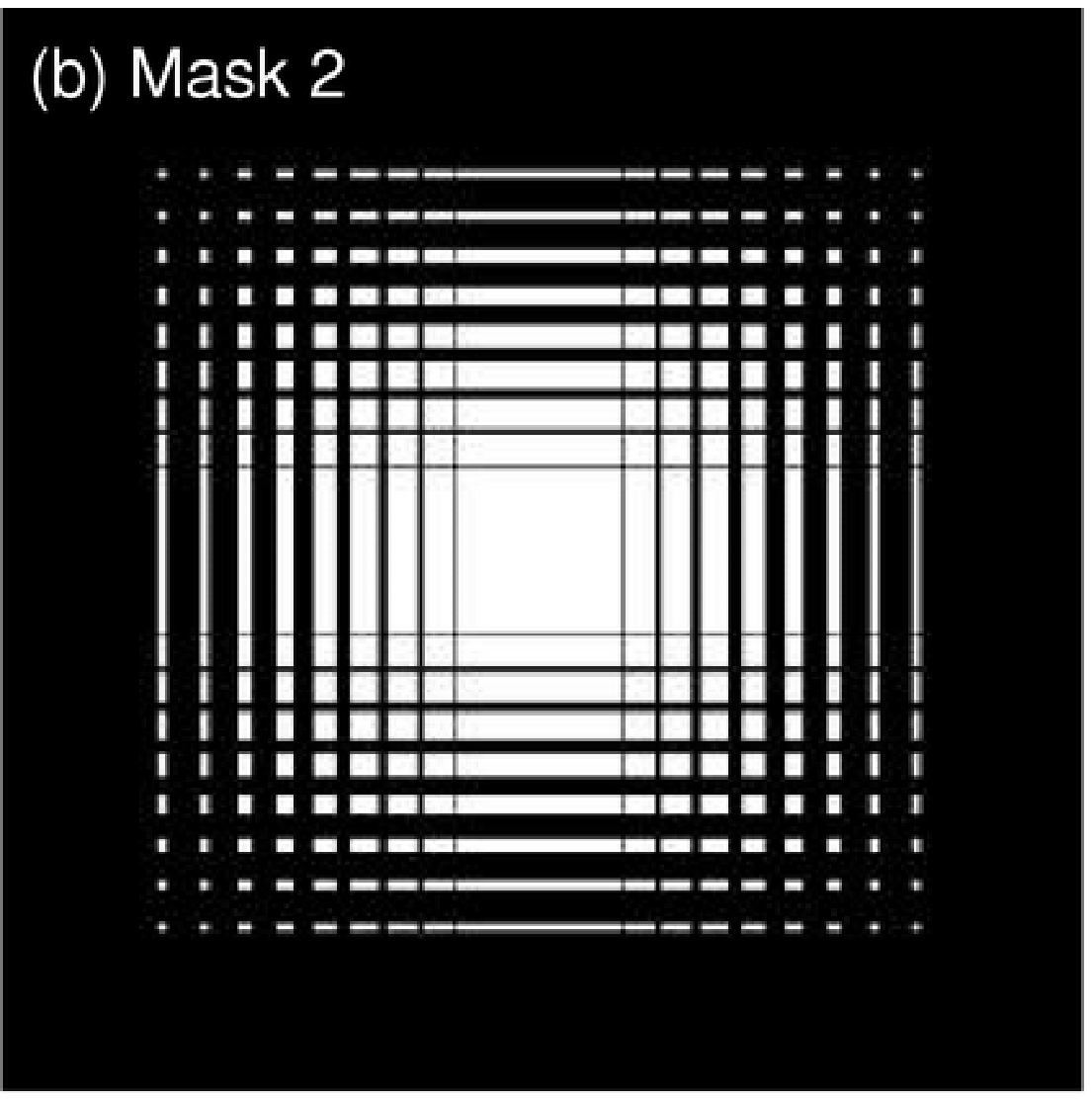}\\
  \includegraphics[width=43mm,height=43mm]{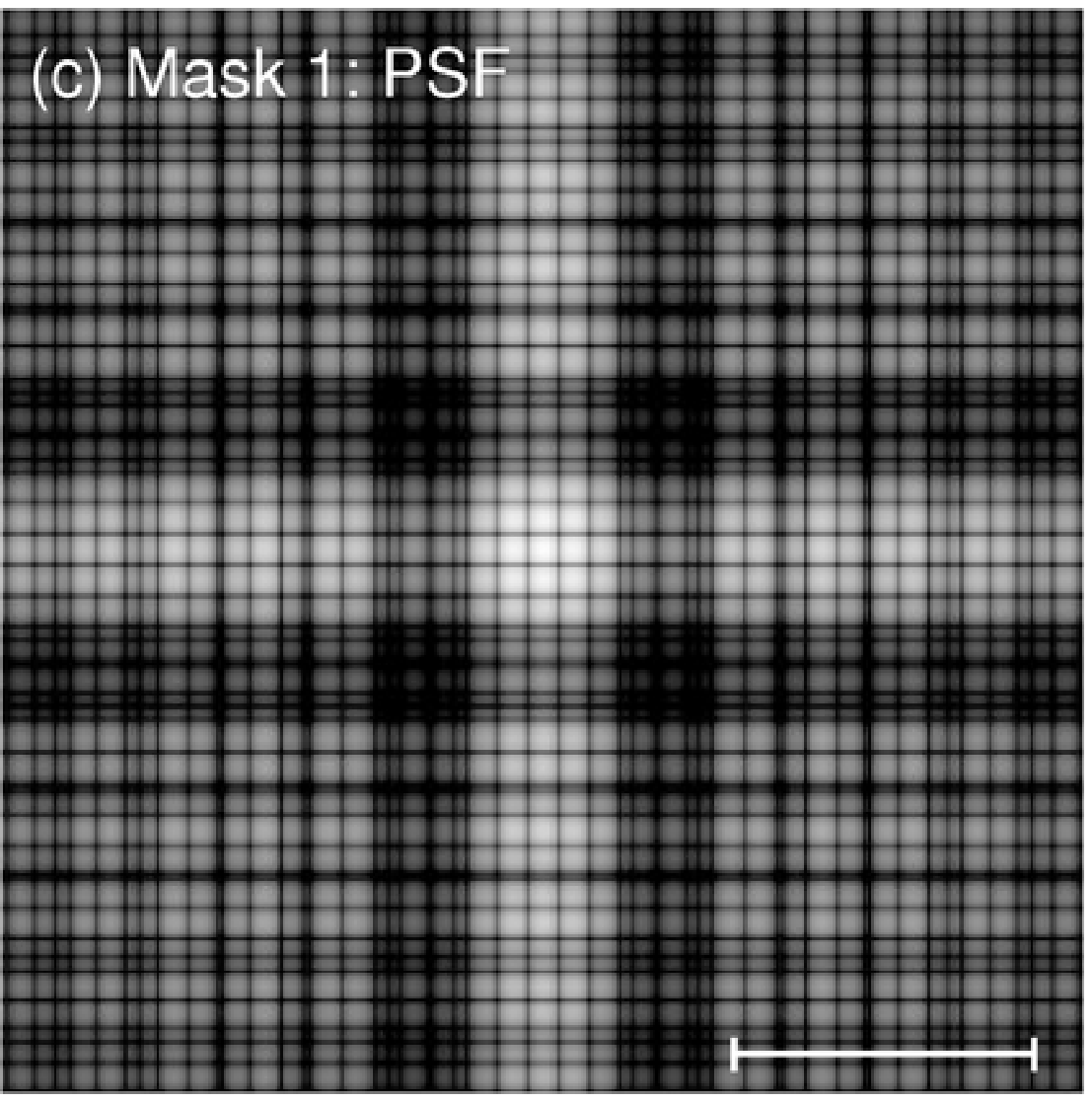}
  \includegraphics[width=43mm,height=43mm]{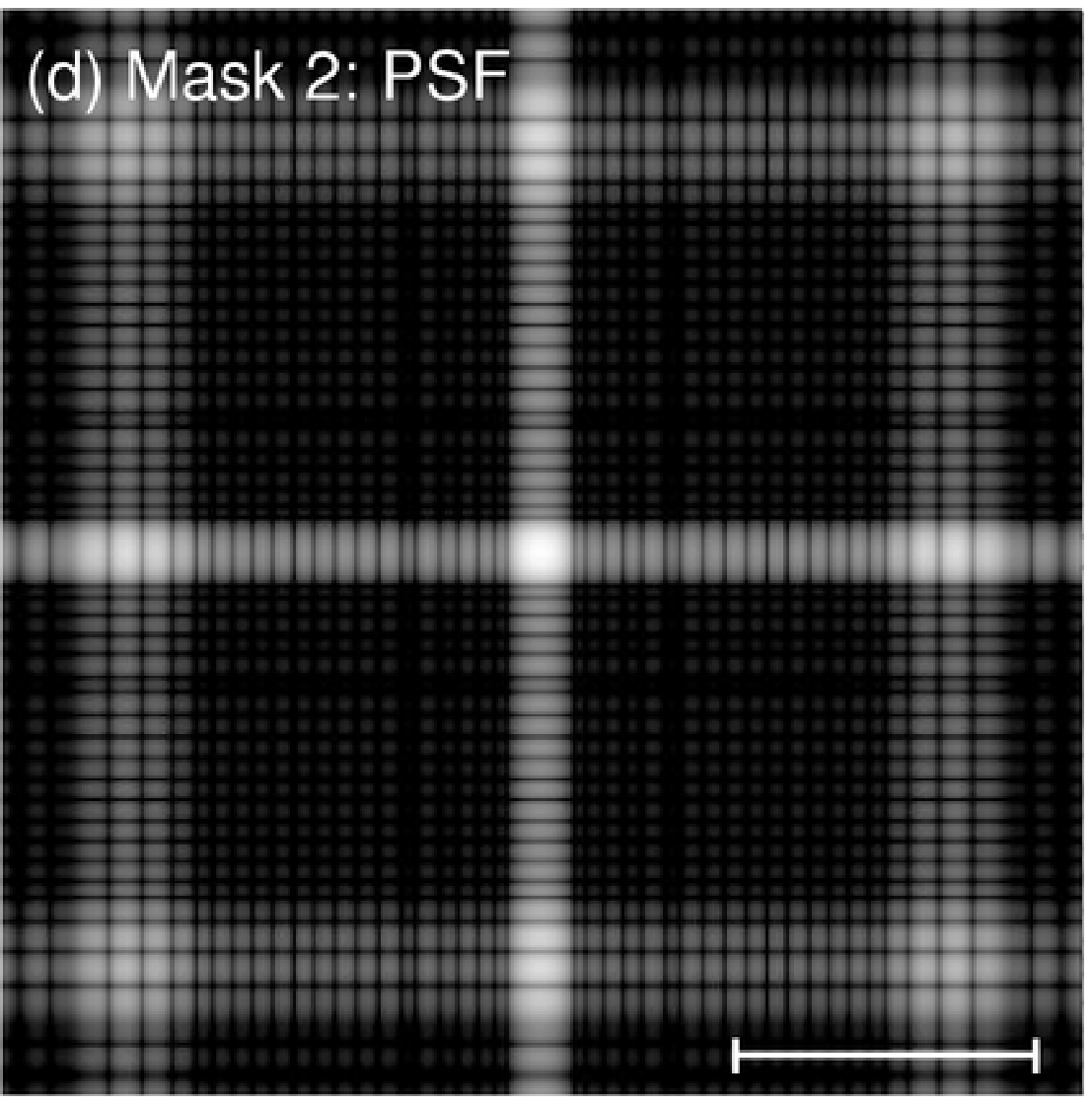}
  \caption{Panels (a) and (b) show the Mask 1
and Mask 2 designs. The transmission through the black and white regions is 0
and 1, respectively . The diameter of the circumscribed circle
to the transmissive part is 2\,mm. Panels (c) and (d) show the
expected (theoretical) PSFs for Masks 1 and 2. The respective
diagonal profiles are shown in Fig.\,\ref{fig05}.
The scale bar is  20\,$\lambda / D$.
   }\label{fig01}

\end{figure}

\begin{figure}[b]
\centering
  \includegraphics[width=80mm]{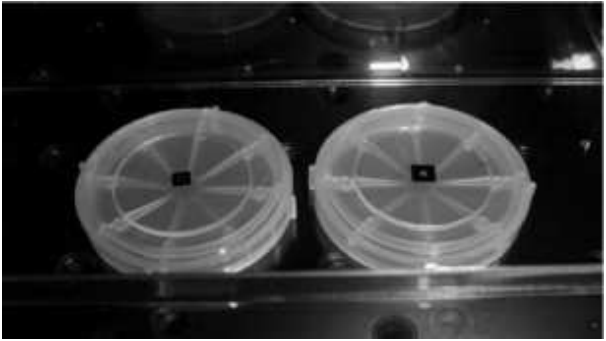}
  \caption{ Photographs showing the fabricated Mask 1 (left) and
Mask 2 (right) on BK 7 substrates. The diameter and
thickness of the substrates is 30.0\,mm and 2.0\,mm, respectively.
The translucent cases beneath the substrates
are used to transport the devices.}\label{fig02}
\end{figure}

In the optimized solutions providing throughput
higher than 15\,\%, the design which had the smallest IWA
was adopted for manufacture. The targeted contrast to
be demonstrated in this work was $10^{-6}$, so that the mask
was designed to achieve a contrast of $10^{-7}$.
The size chosen for the central obstruction of the first mask
(Mask\,1, IWA of 7 $\lambda/D$)
was 30\,\% of the entrance
pupil diameter. 
In our experiment, the wavelength was $\lambda$\,=\,632.8\,nm 
and the diameter of the entrance pupil $D$\,=2\,mm
(i.e. the diagonal of our square masks).
Subsequently, for comparison, a second mask (Mask\,2)
was designed for a pupil without obstruction, and to probe the
coronagraphic sensitivity at much lower IWA (3 $\lambda/D$).
The OWAs for Masks\,1 and 2 were assumed to be 16 and 30\,$\lambda/D$.
The smaller OWA gave a
solution where the rectangular patterns of the mask were
not too narrow and therefore made the fabrication simple
and robust. The larger OWA provided a wider dark region
while it made the fabrication more complex.
Table\,{\ref{table01}} summarizes the specifications
of the masks.
Figure\,{\ref{fig01}} shows the
shape of the masks, and the expected coronagraphic point
spread function (PSF) for Mask\,1 and  Mask\,2.

Each checkerboard mask consists of an aluminum film on a BK7 substrate, and were manufactured using nano-fabrication
technology at the National Institute of Advanced Industrial Science and Technology (AIST) in Japan. A 100\,nm thick
film aluminum was evaporated onto the surface of the substrate. After evaporation, electron beam patterning and a
lift-off process were applied to produce a rectangular transmissive area for the checkerboard pattern. The substrate
was 2\,mm thick and had no wedge angle. The flatness, parallelism and scratch-and-dig of the substrate were less than
$\lambda/10$, 5 arcsecond and $10^{-5}$ respectively. A narrow-band multiple-layer coating was applied to both sides of
the substrate to reduce reflection at the surface after fabrication. The temperature of the mask during coating was
kept less than $100^\circ$\,C to avoid deterioration of the residual resist around the mask pattern. The reduction in
reflectivity of the BK7 substrate was confirmed to be less than 0.25\,\% as a result of this coating.
Figure\,{\ref{fig02}} shows photographs of the final masks on their substrates. Small defects were found on Mask\,2,
but we found no significant degradation in performance for this level of contrast (Sect.\,\ref{section03}).

\subsection{Optical setup}

\begin{figure}
\centering
  \includegraphics[width=90mm]{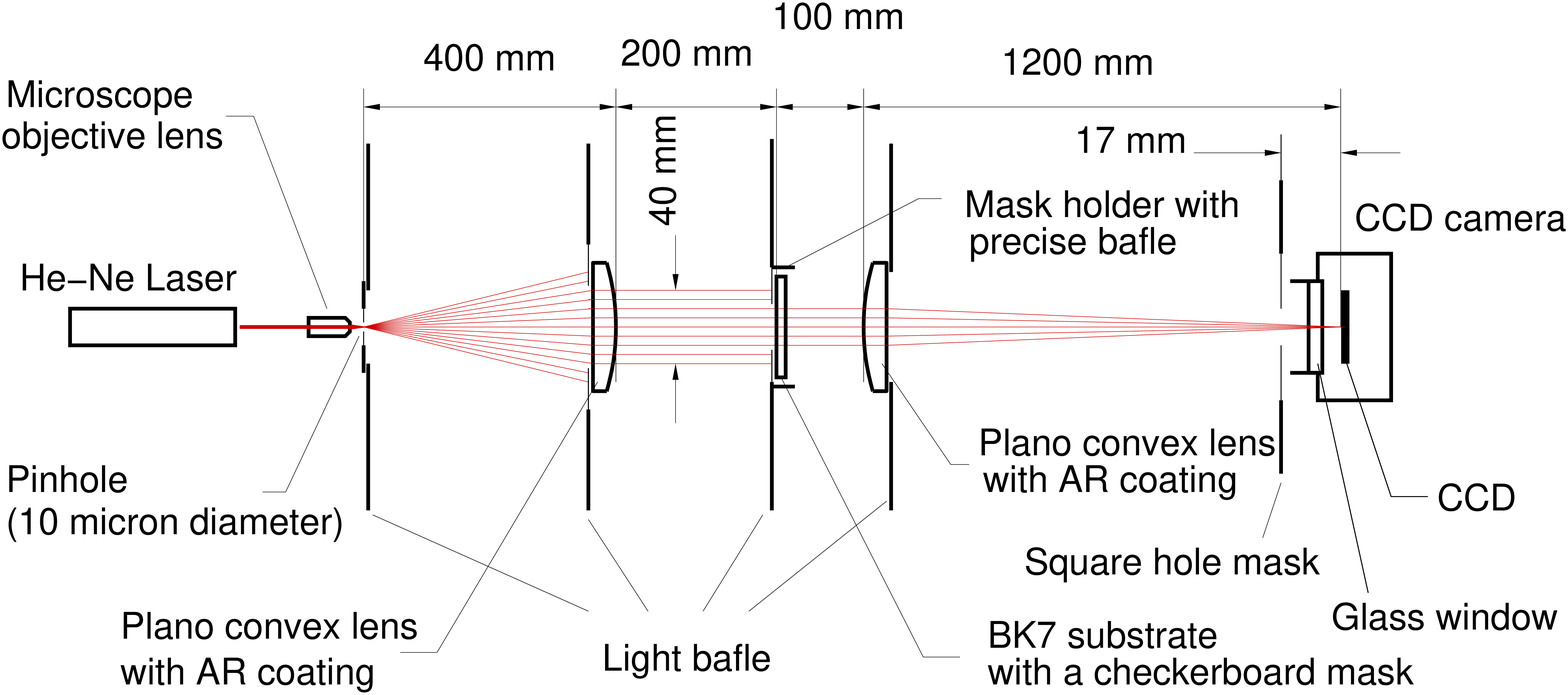}
  \caption{Optical configuration for the experiment. All the
optical devices were set on a table with air suspension in a dark
room with an air cleaning system flowing from the top of the room
onto the optical table. The air flow was either turned on or off to
check its influence on the coronagraphic performance (see text
for details).}
   \label{fig03}
\end{figure}

\begin{figure}
\centering
  \includegraphics[width=90mm]{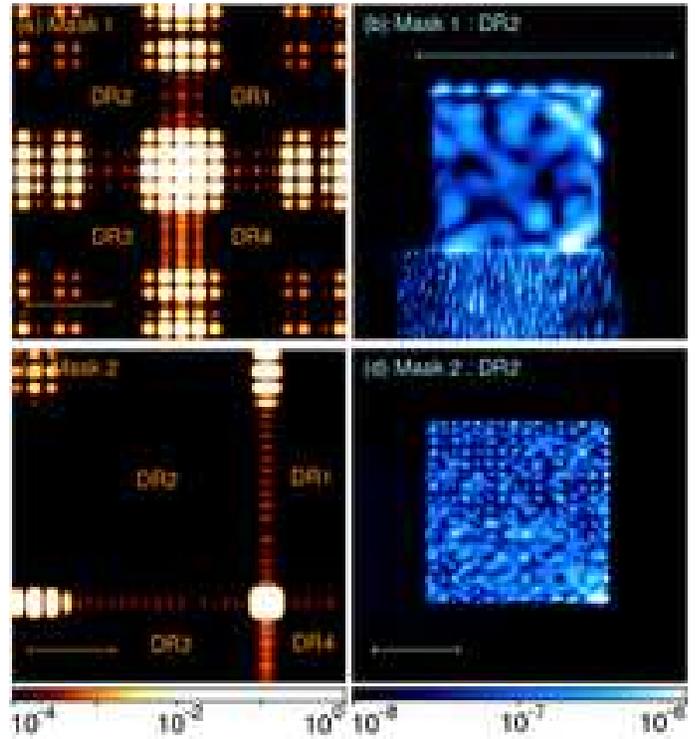}
  \caption{Observed coronagraphic images. Panels (a) and (c)
show images including the core of the PSF for Masks 1 and 2. The tail below the bright peak is due to a readout effect.
Panels (b) and (d) are images of the dark region obtained with a mask with a square aperture. The prickle like pattern
below the square aperture in (b) is the result of reflection by the support structure of the aperture for Mask 1. The
scale bar is  10\,$\lambda / D$.
   }
  \label{fig04}
\end{figure}

The configuration of the experiment is shown
in Fig.\,\ref{fig03}.
All the optics were set in a dark room with an air cleaning
system. Clean air flowed from the top of the room
to the optical table during the settings and measurements.
Although we plan to use an adaptive optics system in
future experiments, it was not used in this work.
A He-Ne laser was used as the light source. A spatial
filter consisting of a microscope objective lens and a 10\,$\mu$m
diameter pinhole was also used.

A 50\,mm diameter BK7 plano-convex lens was used for collimating the entrance beam. The lens is a commercially
available one, and the surface quality was better than  2\,$\lambda$ (over a 30\,mm diameter). The quality of the
collimation was checked and adjusted by a Hartman test. The error in the residual wavefront caused by alignment errors
in the collimation lens was in ${\pm}\,\lambda/600$. A usual narrow-band multiple-layer anti-reflection coating was applied
to both sides of the lens to reduce reflection at the surface. 
The residual reflectivity was confirmed to be less than
0.25\,\%. A light baffle with an edge thickness of 50\,$\mu$m was used to produce a collimated beam of 40\,mm diameter.

The pupil mask was fixed in a holder with a precisely
manufactured baffle with a circular hole made by electrical
discharge machining. The thickness of this baffle was  20\,$\mu$m
and the diameter of the hole was 3.5\,mm. A 20\,$\mu$m thick spacer
was used between this mask and the pupil
mask. A black coating was applied to both sides of the baffle
and the spacer.

Another plano-convex lens was used for focusing. The
focal length of this lens was 1200\,mm and the other specifications
were the same as those of the collimating lens.

A commercially available cooled CCD camera (BITRAN) with  2048\,${\times}$\,2048 pixels was used to measure the PSF. The
overall chip size was 15.16\,mm\,${\times}$\,15.16\,mm with a flat glass window in front. The CCD was cooled and stabilized at
0$^\circ$\,C throughout the experiment. The camera was mounted on a linear motor drive stage in order to scan along the
optical axis to find the best focus position.

If the mask is well aligned to the optical axis,
reflected light (due to reflection from the mask itself) produces
a very well focused point on, or very close to, the
pinhole substrate, which is generally reflective.
To prevent such reflection, whole of the pupil mask
holder with the mask
and the thin baffle was tilted 0.5 degree
from the optical axis in the direction of the diagonal of the mask.
This ensured that the backward focused spot was far away from
any potentially reflective surface, and thus deviated sufficiently
from the optical axis. The CCD camera was also tilted
by 0.5 degree but in the opposite direction to the
mask in order to avoid multiple reflections between these two
components. Other devices were set perpendicular
to the optical axis. The angular accuracy of the setup was
$\sim\,0.1$\,degree and the accuracy of the centering was
$\sim\,0.2$\,mm.

\subsection{Measurement}

The dynamic range of the CCD was insufficient to
capture the full dynamic range of the PSF in one single exposure.
Therefore, several exposure times were used: 0.03,
0.1, 1.0, 10, 100, and 1800 seconds. The 1800 seconds
exposure was for measurements in the dark region. The
brightest region of the PSF saturated the detector even
with the shortest exposure time, so we used additional
neutral density filters with an optical density of
2. We also carefully checked the CCD linearity by imaging
the PSF core through a combination of neutral density
filters with a total optical density of 6, and with a 1800
second exposure. We found that no significant correction
was needed so that we could rely on the multi-exposure
time method.

For measurements of the dark region, we used
a square-shaped mask located just in front of the camera window
to block the flux outside the dark region.
Without this, the contrast did not exceed $10^{-5}$ because of
inner scattering or multi-reflections within the
camera surfaces. The role of this mask is equivalent
to the ``bow tie mask'' shown
in Belikov et al.\,(2006),  Kasdin et al.\,(2005),
though our mask was set farther
in front of the camera window ($\sim$17\,mm).
The size of the
square aperture was 2.3\,mm\,${\times}$\,2.3\,mm for pupil Mask 1 and
8.0\,mm\,${\times}$\,8.0\,mm for pupil Mask 2 with the
corners of the aperture having a radius of
0.2\,mm.
The thickness of
the mask was 20\,$\mu$m  and a black coating was applied to both
sides. The mask was set so that the corners of the aperture were
7.5\,$\lambda/D$ and 3.5\,$\lambda/D$
distant from the core of the PSF for the experiments with Mask 1
and Mask 2, respectively.

After every exposure for the PSF measurement,
we blocked the beam with a black plate placed close to
the CCD side of the pupil mask, and took a dark image with the same
exposure time. Therefore, this image includes the influence of
both the effect of the dark current of the CCD and the background light.
This dark frame was subtracted from the corresponding
PSF image.


\section{Result and Discussion}\label{section03}

\begin{figure}
\centering
  \includegraphics[width=90mm, height=70mm]{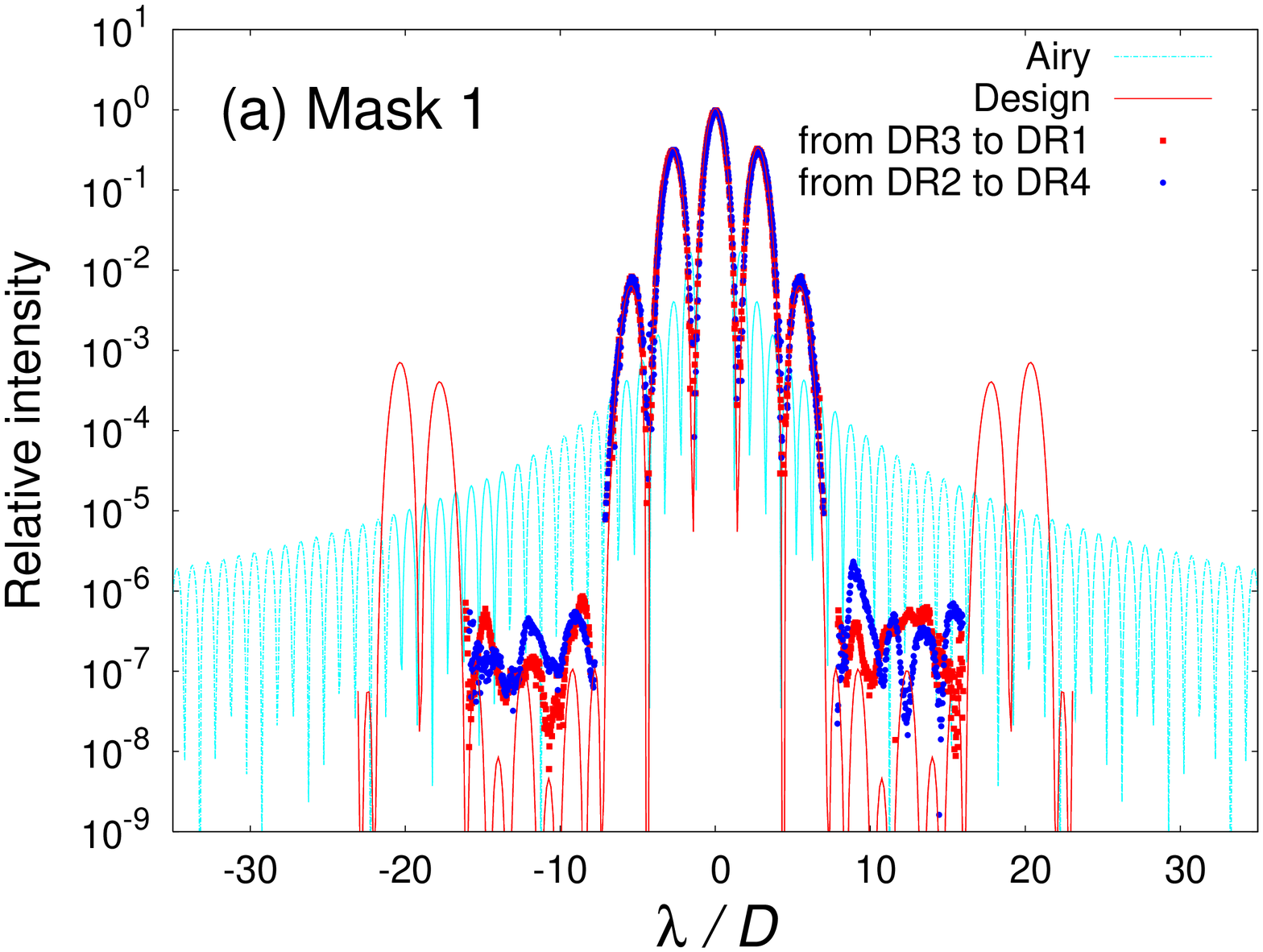}
  \includegraphics[width=90mm, height=70mm]{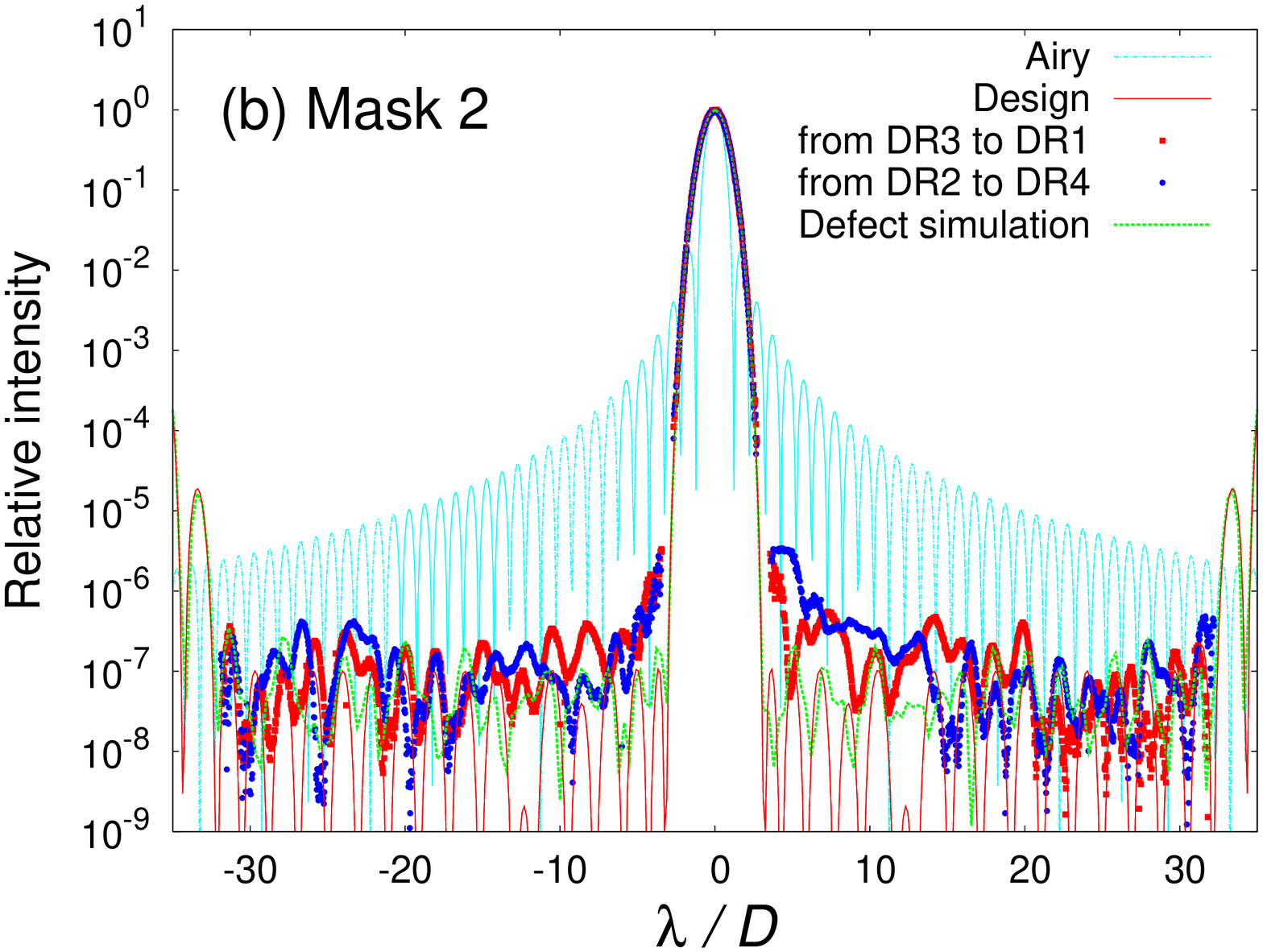}
  \caption{ Diagonal profiles of the coronagraphic image for Mask 1
(a) and Mask 2 (b). Observed and theoretical profiles are shown
as well as the theoretical Airy profile. Each profile is
normalized by the peak intensity in each image. For Mask 2,
the result of simulation with the observed mask defects shown
in Fig. {\ref{fig06}} is also presented.
   }\label{fig05}
\end{figure}


Observed coronagraphic images and their profiles with Masks\,1 and Masks\,2 are shown in Fig.\,{\ref{fig04}} and
Fig.\,{\ref{fig05}}. For both masks, the profiles of the core of the PSF are quite consistent with those expected from
theory. The majority of the area of the dark region in the PSF with Mask 1 is less than 10$^{-6}$ as shown in
Fig.\,{\ref{fig05}}. On average, the observed contrast for each dark region on a linear scale is $3.4 {\times} 10^{-7}$, $2.1
{\times} 10^{-7}$, $1.8 {\times} 10^{-7}$ and $3.5 {\times} 10^{-7}$ for DR\,1, DR\,2, DR\,3 and DR\,4 respectively, where DR 1  $\sim$ DR 4
are the dark regions corresponding to the quadrants around the core shown in Fig.\,{\ref{fig04}}. The average contrast
for all the dark regions is $2.7 {\times} 10^{-7}$. The 3\,$\sigma$ detection limit computed within each observed dark region
DR\,1 $\sim$ DR\,4 is $7.5 {\times} 10^{-7}$, $5.8 {\times} 10^{-7}$, $4.1 {\times} 10^{-7}$, and $9.1 {\times} 10^{-7}$, respectively. PSF
subtraction or fitting has the potential to provide better performance, though such methods require assumptions about
the repeatability of the pattern observed in the dark regions or the shape of the PSF. Therefore, the simple
3\,$\sigma$ limit corresponds to a conservative estimate of the detection limit. Nevertheless, all the values are below
our 10$^{-6}$ limit. 
Considering the effective IWA (including local bright speckles contributions), we
can define for each quadrant the distance beyond which the contrast falls below 10$^{-6}$. If we call this distance
${IWA}_{6}$, it is less than $7.5\,\lambda/D$ for DR\,1, DR\,2 and DR\,3, while it is $9.5\,\lambda/D$ for DR\,4. These
results suggest that the average of ${IWA}_{6}$ is smaller than $8.0\,\lambda/D$ and is quite close to theoretical
value of 7.0\,$\lambda/D$.

On the other hand, Fig.\,{\ref{fig04}}\,(b)
exhibits an irregular flux distribution in the dark region,
which is not predicted by the
theoretical PSF of the mask.
Consequently the obtained contrast
does not reach the theoretical 10${^{-7}}$ value.
It was confirmed that intensity distribution of
the irregular pattern was repeatable
in a fixed  setup configuration.
We did the same measurement with and
without air flow or air suspension
of the optical table to test the influence of air turbulence
or vibration of the devices, and confirmed no significant
change occurred.
On the contrary, rotating or shifting the
mask parallel to its surface changed the shape and intensity
of the irregular pattern. Hence, we concluded that
errors in the mask fabrication do not dominate the
coronagraphic performance of Mask 1.
Brighter speckles are caused by a combination
of effects in the beam-line:
wavefront errors, multi-reflections and scattering by microscopic
defects on the surface of the optics and can therefore
be reasonably considered as being limiting factors in this
experiment.


For the coronagraphic image with Mask\,2, the average contrast of DR\,1, DR\,2, DR\,3 and DR\,4 on a linear scale is
$1.1 {\times} 10^{-7}$, $1.0 {\times} 10^{-7}$, $9.0 {\times} 10^{-8}$  and $1.3 {\times} 10^{-7}$, respectively. The average contrast of the whole
dark region is $1.1 {\times} 10^{-7}$. 
${IWA}_{6}$ for DR\,1 $\sim$ DR\,4 is  $4.2\,\lambda/D$,
$4.1\,\lambda/D$, $4.5\,\lambda/D$, $5.5\,\lambda/D$, respectively. The
average of $IWA_6$ is $4.6\,\lambda/D$. 
The 3\,$\sigma$  limit is $3.8 {\times} 10^{-7}$, $3.0 {\times}10^{-7}$,  $2.7 {\times} 10^{-7}$ and $3.8 {\times} 10^{-7}$ for
DR\,1, DR\,2, DR\,3 and DR\,4 respectively. In the bottom part of Fig.\,{\ref{fig04}}\,(d), an irregular speckle
pattern can be observed, getting brighter close to the optical axis (bottom right). This pattern was invariant in
repeated measurements with a fixed setup. Change in the air flow and suspension conditions made no significant
difference in the observed image. On the contrary, when we shifted and rotated the mask, this speckle pattern changed,
suggesting the same limiting factors as for Mask 1.

On the other hand, a lattice-like pattern was also found
at a relatively large distance from the core in the dark region
with Mask 2. The position of each lattice ``node'' agrees
quite well with the ones expected from the theoretical
image as shown in Fig. {\ref{fig02}},
but their intensity is not uniform
and slightly brighter than that predicted by theory.
However, we noticed that
their intensity changed while shifting and/or rotating
the mask relative to the beam line optics.
We therefore concluded that we were observing the
theoretical pattern
modulated by speckles originating from aberrations
in the optics.

We attempted to simulate the effects of the observed defects,
in order to check if the recorded intensity of the slightly
brighter (than theory) lattice pattern was due to these
defects.
As shown in Fig.\,{\ref{fig06}}\,(a) and (b),
two defects were found
in Mask 2 by careful inspection of microscopic images.
The simulated image including these defects is shown in
Fig.\,{\ref{fig06}}\,(c),
and the corresponding diagonal cut appears in
Fig.\,{\ref{fig05}}\,(b).
To reproduce the observed intensity level of the
lattice nodes, a slight overestimate of the size of the defects
was needed in the simulation. As a result, we obviously
failed to reproduce the observed image because the lattice
pattern seen in the top part of
Fig.\,\ref{fig04}\,(d) is much more
similar to the theoretical pattern than to the simulated image
with defects. We can therefore confidently conclude that
these defects do not significantly degrade
the coronagraphic performance at this level of contrast.

Belikov et al.\,(2006),  Kasdin et al.\,(2005)
presented some profiles of coronagraphic PSF using a free
standing shaped pupil mask. The performance achieved by our Mask\,2 is comparable, or somewhat higher than the
previously mentioned results. But the direct comparison must be balanced because the aim and strategy were different.
In our experiment, the beam size was quite smaller ($\sim 1/10$), which is definitely an advantage to reduce the wave
front error caused by the surface figure of optical components. On the other hand, the centimeter-size beam from, e.g.
Belikov et al.\,(2006)
was suitable for using a deformable mirror where the aim was also to test some focal plane speckle
attenuation. On-substrate, transmissive components we used can be regarded as being disadvantageous because of the
potential multi-reflections and diffusion effects.

As shown above, it has been demonstrated that both
checkerboard masks, with or without a large obstruction
in the pupil work well, and the observed performance
exceeds the target contrast of 10$^{-6}$.
In principle,
it is possible to construct a coronagraph with self standing
checkerboard mask and mirrors.
Such system is essentially free from
the imperfectness of transmissive devices,
which provide difficulties of development.
The simplicity of
the optics is another advantage of the shaped pupil mask
coronagraph especially for a general purpose telescope
which carries other instruments and therefore
the space
and weight assigned for the coronagraphic instrument is
strictly limited. These facts suggest that a coronagraph using a
checkerboard mask is a very promising solution for
general purpose infrared telescopes with a significantly
obstructed pupil, such as SPICA.
On the other hand, beam blocking by the pupil mask,
reflected flux from the CCD itself, multi-reflections between
the optical devices and inner scattering in the camera can
decrease the contrast.
Band pass filters or other optical
devices required for astronomical purposes are indispensable
for the actual instrument, but they make the design of
the instrument more complex. Demonstration of the
coronagraphic part is therefore not the only critical part to
address, and global optimization of all the other components
is mandatory in order to achieve the expected performance.

\begin{figure}
\centering
  \includegraphics[height=45mm]{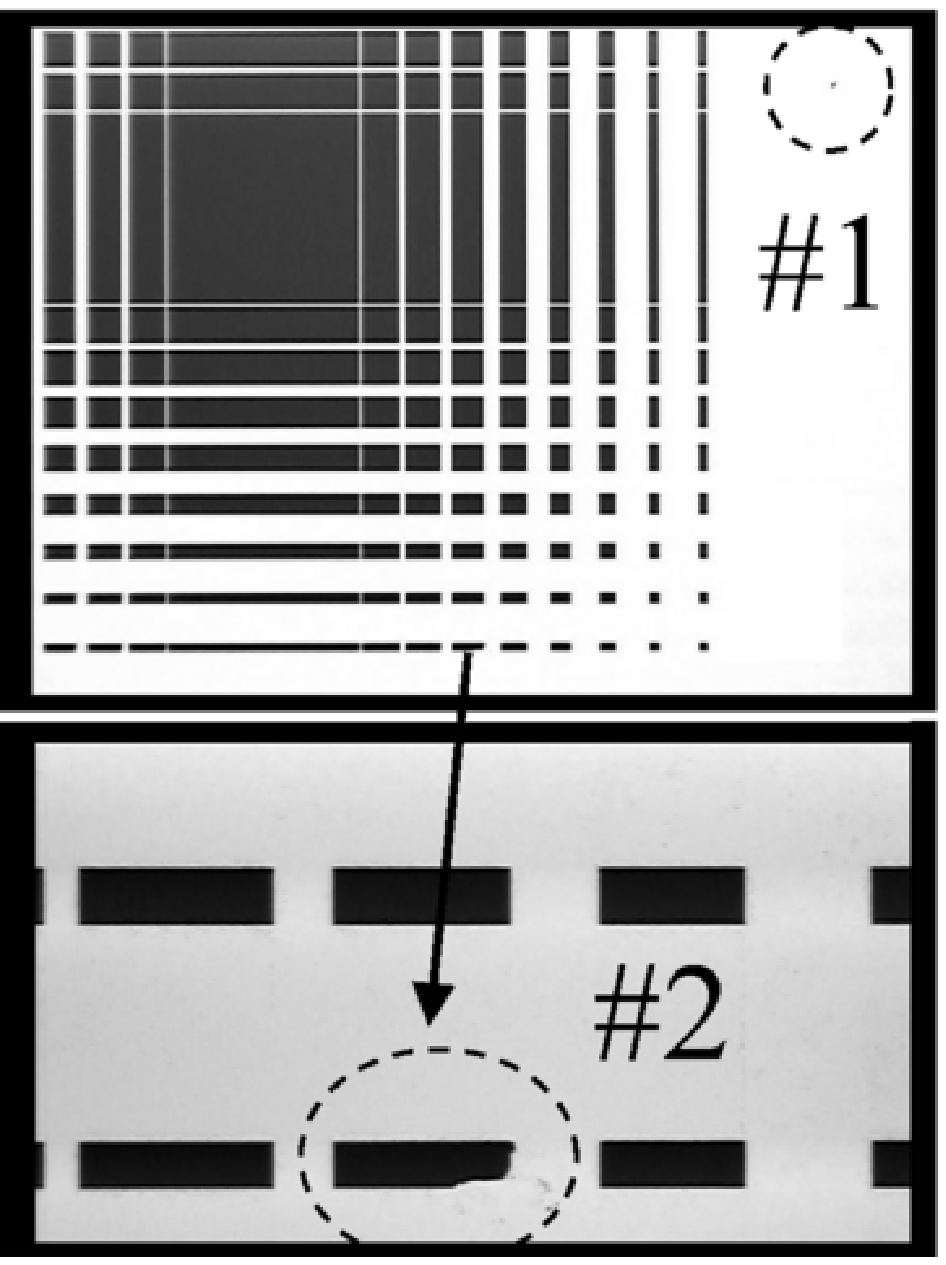}
  \includegraphics[height=45mm]{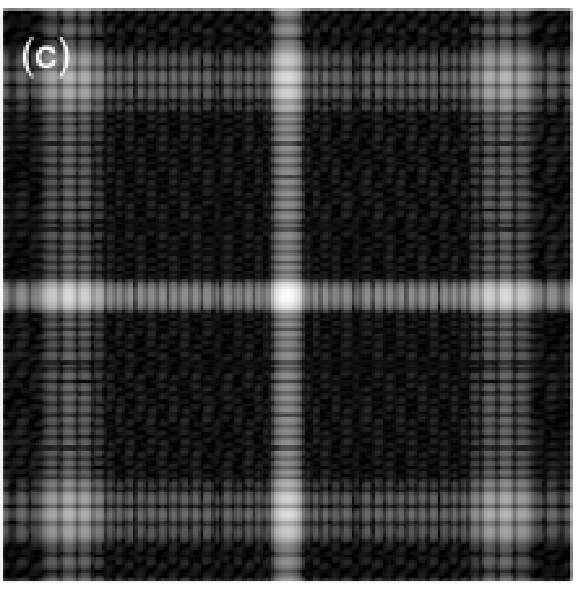}
  \caption{ Panels (a) and (b) show the defects in Mask 2 observed
through an optical microscope using reflected light. The darker
parts in the images correspond to the transmitted areas. Panel (c)
Simulated coronagraphic image including the influence of the
observed defects. The color scale is the same as
in  Fig.\,{\ref{fig01}}.
}\label{fig06}
\end{figure}

\section{Conclusion}

We report the results from a laboratory experiment using a binary
checkerboard-type pupil shaped mask coronagraph.
Although this kind of mask presents several limitations
(large IWA and low throughput), its manufacture
and implementation are very robust. Two masks, optimized for
either obstructed pupil or not, were fabricated
and tested. Both masks achieved a higher performance than
the $10^{-6}$  requirement:
$2.3\,{\times}\,10^{-7}$ and $1.1\,{\times}\,10^{-7}$.
Therefore,
we conclude that a binary checkerboard mask
is a very attractive solution for the future MIR coronagraph
of the SPICA space telescope, for which this study
was carried out. Other coronagraphic solutions, more complex
to implement are also considered, but the checkerboard
mask can be regarded as the safest option.

\begin{acknowledgements}
We are grateful to T. Wakayama, T. Sato,
N. Nakagiri and other colleagues in AIST for their great
support in the mask fabrication. We would like to thank all
persons relating to SPICA and the SPICA coronagraph. This
work was supported in part by a grant from the Japan Science
and Technology Agency. LA is supported by Grants-in-Aid
(No. 160871018002) from the Ministry of Education, Culture,
Sports, Science, and Technology (MEXT) of Japan.
\end{acknowledgements}

\end{document}